\documentclass[prd,aps,superscriptaddress,nofootinbib,showpacs,twocolumn]{revtex4}

\usepackage[intlimits]{amsmath}
\usepackage{amsfonts}
\usepackage{amssymb,amscd}
\usepackage{color}
\usepackage[dvips]{epsfig}

\newcommand{\beq}{\begin{equation}}
\newcommand{\eeq}{\end{equation}}
\newcommand{\beqa}{\begin{eqnarray}}
\newcommand{\eeqa}{\end{eqnarray}}
\newcommand{\be}{\begin{equation}}
\newcommand{\ee}{\end{equation}}

\newcommand{\vp}{\vec{p}}
\newcommand{\vq}{\vec{q}}
\newcommand{\vk}{\vec{k}}
\newcommand{\op}{\omega_{p}}
\newcommand{\oq}{\omega_{q}}
\newcommand{\ok}{\omega_{k}}

\newcommand{\intk}{\sum_{n_k} \int \frac{d^3k}{(2\pi)^3}}

\def\mA{\Gamma^{A}}
\def\mC{\Gamma^{C}}

\def\eq#1{(\ref{#1})}
\def\Eq#1{Eq.~(\ref{#1})}

\begin{document}

\title{Chiral and deconfinement transition from Dyson-Schwinger equations}
\author{Christian~S.~Fischer}
\affiliation{Institut f\"ur Kernphysik, 
  Technische Universit\"at Darmstadt,
  Schlossgartenstra{\ss}e 9,\\ 
  D-64289 Darmstadt, Germany}
\affiliation{GSI Helmholtzzentrum f\"ur Schwerionenforschung GmbH, 
  Planckstr. 1  D-64291 Darmstadt, Germany.}
\author{Jens~A.~Mueller}
\affiliation{Institut f\"ur Kernphysik, 
  Technische Universit\"at Darmstadt,
  Schlossgartenstra{\ss}e 9,\\ 
  D-64289 Darmstadt, Germany}

\date{\today}
\begin{abstract}
We determine the quark condensate and the dressed
Polyakov loop from the finite temperature Landau gauge
quark propagator evaluated with $U(1)$-valued boundary 
conditions in an approximation to quenched QCD. 
These gauge invariant quantities allow for
an investigation of the chiral and deconfinement 
transition. We compare results from 
Dyson-Schwinger equations on a lattice with infinite 
volume continuum results and study the temperature 
and quark mass dependence of both quantities. In particular
we investigate the chiral condensate and the dressed
Polyakov loop in the chiral limit. We also consider an 
alternative order parameter for the deconfinement 
transition, the dual scalar quark dressing, and compare 
with the dressed Polyakov loop. As a result we find only 
slightly different transition temperatures for the chiral 
and the deconfinement transition at finite quark masses;
in the chiral limit both transitions coincide.

\end{abstract}

\maketitle

\section{Introduction}
The chiral and the deconfinement transition of
QCD are a subject of continuous interest both,
from a theoretical and from an experimental point
of view. Ongoing experiments at the Relativistic 
Heavy Ion Collider (RHIC) and future runs at LHC
and FAIR probe the transition between the high and 
low temperature states of strongly interacting 
quark matter. The transition is characterized by the
breaking and restoration of chiral and center 
symmetry. Hereby it is well
known that both transitions are phase transitions
in the strict sense of the word only in opposite
limits of the theory. In the chiral limit $m=0$ 
chiral quantities such as the chiral condensate 
$\langle \overline{\psi} \psi \rangle$ or the
scalar part $B(0)$ of the inverse quark propagator
at zero momentum are order 
parameters of the chiral phase transition. In the 
absence of dynamical quarks, $m \rightarrow \infty$,
pure gluonic systems are center symmetric in the
low temperature phase. The deconfinement phase transition 
is then related to center symmetry breaking 
indicated by a non vanishing order parameter 
as e.g. the Polyakov loop $L$ \cite{Polyakov:1978vu}.
In the real world, quarks carry (light) masses and 
both transitions are crossovers as nicely established 
by lattice QCD, see e.g. 
\cite{Karsch:2003jg,Aoki:2006br,Bazavov:2009zn,Aoki:2009sc} 
and references therein.

The theoretical search for an underlying mechanism
that links quark confinement and chiral symmetry breaking 
has brought many interesting results in the past years.
Topological field configurations such as center vortices
have been argued to connect both phenomena, see e.g.
\cite{Greensite:2003bk} for a review. In 
\cite{Gattnar:2004gx} a close connection between center
vortices and the spectrum of the Dirac operator has been
revealed; removing those vortices from the gauge field
ensemble resulted in a vanishing string tension and 
a zero chiral condensate via the Casher-Banks relation.

Very recently a number of works 
\cite{Gattringer:2006ci,Bruckmann:2006kx,Synatschke:2007bz,%
Synatschke:2008yt,Bilgici:2008qy,Bruckmann:2008sy,%
Danzer:2008bk,Bilgici:2009tx} have explored the 
relation between chiral and deconfinement signatures
in the spectrum of the Dirac-operator. In 
\cite{Synatschke:2008yt} it has been shown that the 
low lying eigenmodes of the Dirac operator are 
responsible for both, chiral symmetry breaking and
confinement: spectral sums of the Dirac operator may be 
truncated to the lowest lying modes and still lead to 
a linear rising quark-antiquark potential extracted 
from the Polyakov-loop correlator.

One of these spectral sums, the dual condensate or dressed
Polyakov loop, is of particular interest since
it relates the chiral condensate to the Polyakov loop.
This quantity has been studied in lattice gauge theory
\cite{Bilgici:2008qy} and with Dyson-Schwinger equations 
on a torus \cite{Fischer:2009wc} with qualitatively 
similar results. Both of these studies dealt with the 
dressed Polyakov loop on a compact manifold at finite 
quark masses. In this work we add more details to
these investigations, in particular considering the
infinite volume limit, the chiral limit and the large
temperature behavior of these quantities. We also 
investigate another order parameter for the deconfinement
transition, the dual scalar quark dressing function.
Compared to Ref.~\cite{Fischer:2009wc} we also modified
the renormalization condition of the quark mass function in the torus formulation,
thus eliminating temperature effects in the bare quark mass.
As a result we now obtain almost similar transition 
temperatures for the chiral and the deconfinement 
transition for quark masses roughly corresponding to 
an up-quark and identical transition temperatures in 
the chiral limit.

This work is organized as follows: In section 
\ref{sec:order} we outline our two order parameters 
for the confinement-deconfinement transition. Here 
we consider (i) the dual quark condensate as defined 
in Ref.~\cite{Bilgici:2008qy} and (ii) the dual scalar 
quark dressing function as defined below.
Both quantities share the property that they can be 
extracted from the momentum dependence of the quark 
propagator and are therefore accessible with 
functional methods. In this work we use the 
Dyson-Schwinger equations (DSEs) for the quark 
propagator in Landau gauge to determine these 
order parameters. The temperature dependence of the
propagator is evaluated in the Matsubara formalism.
We employ a truncation scheme for the quark-DSE that uses 
$SU(2)$-lattice results 
for the temperature dependent gluon propagator 
\cite{Cucchieri:2007ta} and a temperature dependent 
ansatz for the quark-gluon vertex as input. This 
scheme is detailed in section \ref{sec:DSE}, where we
also discuss our numerical procedure. In 
section \ref{sec:num} we present our numerical results 
in the infinite volume/continuum formulation
of the DSEs and compare with the results of 
Ref.~\cite{Fischer:2009wc} obtained from a formulation 
of the DSEs on a torus. Our infinite volume approach also 
enables us to explore the chiral limit of the theory. We 
compare results for the order parameters specified in section 
\ref{sec:order} and comment on relations to the chiral 
phase transition. Our results are summarized in the
concluding section \ref{sec:sum}. Throughout the paper 
we work in $SU(2)$-Yang-Mills theory, leaving the $SU(3)$ 
case for future work.  

\section{Order parameters for deconfinement \label{sec:order}}

Quantities that transform non trivial under center symmetry 
and thus qualify as order parameters for deconfinement are in general
not easily accessible by functional methods.
However, there has been  considerable progress on this issue in the past years.
In Ref.~\cite{Braun:2007bx}, the ghost and gluon propagators 
of Landau gauge Yang-Mills theory have been used to determine
the Polyakov loop potential within an effective action 
approach. In this work we will focus on an alternative strategy
that allows to extract chiral and deconfinement order parameters 
from a (generalized) quark propagator. This approach establishes links
between dynamical chiral symmetry breaking and quark confinement in QCD.
\cite{Gattringer:2006ci,Bruckmann:2006kx,Synatschke:2007bz,%
Bilgici:2008qy,Fischer:2009wc,heid}.

\subsection{The dual condensate \label{dualcond}}

The dual quark condensate as order parameter for center symmetry 
has emerged from a series
of works connecting spectral sums of the Dirac propagator with 
Polyakov loops and their correlators 
\cite{Gattringer:2006ci,Bruckmann:2006kx,Synatschke:2007bz}. 
Within the framework of lattice gauge theory 
the dual condensate $\Sigma_n$ has been introduced in Ref.~\cite{Bilgici:2008qy}. 
It is defined by the phase-Fourier-transform
\beq \label{dual}
\Sigma_n = \int_0^{2\pi} \, \frac{d \varphi}{2\pi} \, e^{-i\varphi n}\, 
\langle \overline{\psi} \psi \rangle_\varphi
\eeq
of the ordinary quark condensate evaluated with respect to a phase 
$e^{i\varphi}$ in the temporal direction of the Euclidean theory. 
This phase is introduced by the generalized, $U(1)$-valued boundary 
condition $\psi(\vec{x},1/T) = e^{i \varphi} \psi(\vec{x},0)$ in the 
temporal direction. For the usual antiperiodic boundary conditions for 
fermions we have $\varphi=\pi$, whereas $\varphi=0$
corresponds to periodic boundary conditions. The $\varphi$-dependent
quark condensate $\langle \overline{\psi} \psi \rangle_\varphi$ is
proportional to a sum over closed loops winding $n$-times around 
the compact time direction: 
\beq \label{loop}
\langle \overline{\psi} \psi \rangle_\varphi =  
\sum_{l \in \mathcal{L}} \frac{e^{i\varphi n(l)}}{(a m)^{|l|}} U(l) \,.
\eeq
Here $\mathcal{L}$ denotes the set of all closed loops $l$ with
length $|l|$ on a lattice with lattice spacing $a$. 
Furthermore $m$ is the quark mass.  
$U(l)$ stands for the chain of link variables in a loop $l$
multiplied with appropriate sign and normalization factors, 
see Ref.~\cite{Bilgici:2008qy} for details. Each loop
that closes around the temporal boundary picks up factors of 
$e^{\pm i\varphi}$ according to its winding number $n(l)$.
The Fourier transform in \eq{dual}
projects out those loops with winding number $n$. The
dual condensate $\Sigma_1$ then corresponds to loops that wind 
exactly once and is called the 'dressed Polyakov loop' \cite{Bilgici:2008qy}.
This quantity transforms under center transformation identically
as the conventional Polyakov loop \cite{Polyakov:1978vu} 
and is therefore an order parameter for center symmetry. 
The numerical agreement between dressed and conventional 
Polyakov loop has been established for gauge groups $SU(3)$ 
\cite{Bruckmann:2008sy} and, remarkably, also for the 
centerless $G(2)$ \cite{Danzer:2008bk}. 

These lattice calculations have been carried out in quenched QCD
and in a formulation where the nontrivial $U(1)$-valued boundary
conditions are introduced for the measured operators only. Such 
a formulation avoids the complications associated with the 
introduction of nontrivial boundary conditions into the generating
functional of QCD with fermions. The latter option is formally 
equivalent to the introduction of an imaginary chemical potential 
leading to the pattern of Roberge-Weiss periodicity of
the generating functional \cite{Roberge:1986mm}. In this work we
will approximate the quenched lattice formulation and therefore
do not encounter the Roberge-Weiss symmetry. A different
approach is followed, however, in the (unquenched) renormalization 
group framework of Ref.~\cite{heid}, where the Roberge-Weiss symmetry
is explicitly taken into account.

Furthermore, a comment on regularization is in order here.
Below we will determine the dual condensate $\Sigma_1$ from 
Dyson-Schwinger equations in the infinite volume and continuum 
limit. For non-vanishing quark masses the quark condensate is
quadratically divergent and needs to be properly regularized, 
see e.g. Ref.~\cite{Williams:2006vva} for a recent discussion. 
Correspondingly, the loop expansion \Eq{loop} breaks down for 
$a \rightarrow 0$. In this work we will employ an ultraviolet 
regulator in the form of a simple cutoff. For large enough
cutoffs the regulator dependent part of the condensate is 
independent of temperature and therefore does not affect the 
chiral transition temperature. The regulator dependent 
part of the condensate is also independent of the boundary 
angle. Consequently it does not appear in the dual condensate 
and does not affect the deconfinement transition temperature 
$T_{dec}$ either. The latter property can also be demonstrated
{\it a posteriori} by comparing $T_{dec}$ from the dual condensate 
with the one obtained from the dual scalar quark dressing function 
introduced in the next subsection. Since the (dual) scalar quark 
dressing function does not suffer from regularization problems, 
agreement between the corresponding transition temperatures
indicates that our regularization procedure is adequate. This is 
indeed the case as demonstrated in section \ref{sec:resdual}. 

\subsection{The dual scalar quark dressing function}

Another order parameter for the deconfinement transition is the 
phase-Fourier-transform of the scalar quark dressing function 
evaluated at lowest Matsubara frequency and zero momentum. To
derive this quantity note first that in 
momentum space the $U(1)$-valued boundary conditions 
introduced above result in Matsubara modes 
$\op(n_t,\varphi) = (2\pi T)(n_t+\varphi/2\pi)$ in the 
$p_4$-direction, which depend on the boundary angle 
$\varphi \in [0,2\pi[$. The inverse quark 
propagator can then be written as  
\beq \label{quark}
S^{-1}(\vp,\op) = i \gamma_4\, \op C(\vp,\op) 
+ i \gamma_i \, p_i A(\vp,\op) + B(\vp,\op) \,,
\eeq
with vector and scalar quark dressing functions $C,A,B$. A further
tensor component proportional to $\sigma_{\mu \nu}$ is possible
in principle but can be omitted in all practical calculations 
\cite{Roberts:2000aa}. 
\begin{figure}[t]
\centerline{\includegraphics[width=\columnwidth]{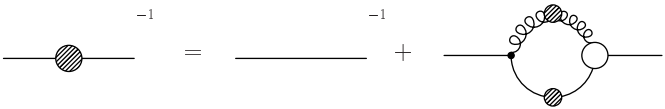}}
\caption{The Dyson-Schwinger equation for the quark propagator. Filled circles
denote dressed propagators whereas the empty circle stands for the dressed
quark-gluon vertex.}
\label{fig:quarkDSE}
\end{figure}

The scalar quark dressing function $B(\vp,\op)$ evaluated at $\vp=0$ 
and $\op(0,\pi) = \pi T$ is an order parameter for the chiral phase 
transition in the chiral limit. It also serves as indicator for the chiral 
crossover at non-vanishing bare quark masses in a similar way as the 
chiral condensate does. Its dual
\beq 
\Sigma_B = \int_0^{2\pi} \, \frac{d \varphi}{2\pi} \, e^{-i\varphi}\, 
B(0,\op(0,\varphi))\,,
\eeq
is an order parameter for the confinement-deconfinement
transition. This can be shown in a similar fashion as for the
spectral sums discussed in \cite{Synatschke:2008yt}:
Using $B(0,\op(0,\varphi)) = 1/4 \,\mbox{tr} [S^{-1}(0,\op(0,\varphi))]$ 
we have
\beqa 
\Sigma_B = \int_0^{2\pi} \, \frac{d \varphi}{8\pi} \, e^{-i\varphi}\, 
\int d^3x \, \int_0^{1/T} dx_0 \, \mbox{tr} 
\langle \vec{x},x_0 | D_{\varphi}^{-1} | 0 \rangle^{-1} \label{dualskalar}
\eeqa
where $D_{\varphi}$ is the (massive or chiral) Dirac operator evaluated
under presence of the $U(1)$-valued boundary conditions. A center 
transformation on $\Sigma_B$ introduces an additional phase factor
$z = e^{i 2 \pi k/N}$ with $k=0,\dots,N_c-1$ for $N_c$ colors, which adds
to the phase $e^{i\varphi}$ for our $U(1)$-valued boundary conditions.
We then obtain
\beqa 
^z\Sigma_B &=& \int_0^{2\pi}  \!\frac{d \varphi}{8\pi}  e^{-i\varphi} 
\!\!\int d^3x \!\!\int_0^{1/T} dx_0 \mbox{tr} 
\langle \vec{x},x_0 | D_{\varphi+2\pi k/N}^{-1} | 0 \rangle^{-1}  \nonumber\\
           &=& z\, \int_0^{2\pi} \! \frac{d \varphi}{8\pi}  e^{-i\varphi}\, 
\!\!\int d^3x \!\!\int_0^{1/T} dx_0 \mbox{tr} 
\langle \vec{x},x_0 | D_{\varphi}^{-1} | 0 \rangle^{-1}\,,
\eeqa
i.e. the dual scalar quark dressing $\Sigma_B$ transforms under center
transformations exactly like the conventional Polyakov loop and therefore
acts as order parameter for the deconfinement transition in the heavy quark 
limit. This will be confirmed by our numerical results given below.

\section{Dyson-Schwinger equations for the quark propagator \label{sec:DSE}}

\subsection{The truncation scheme \label{trunc}}

\begin{figure*}
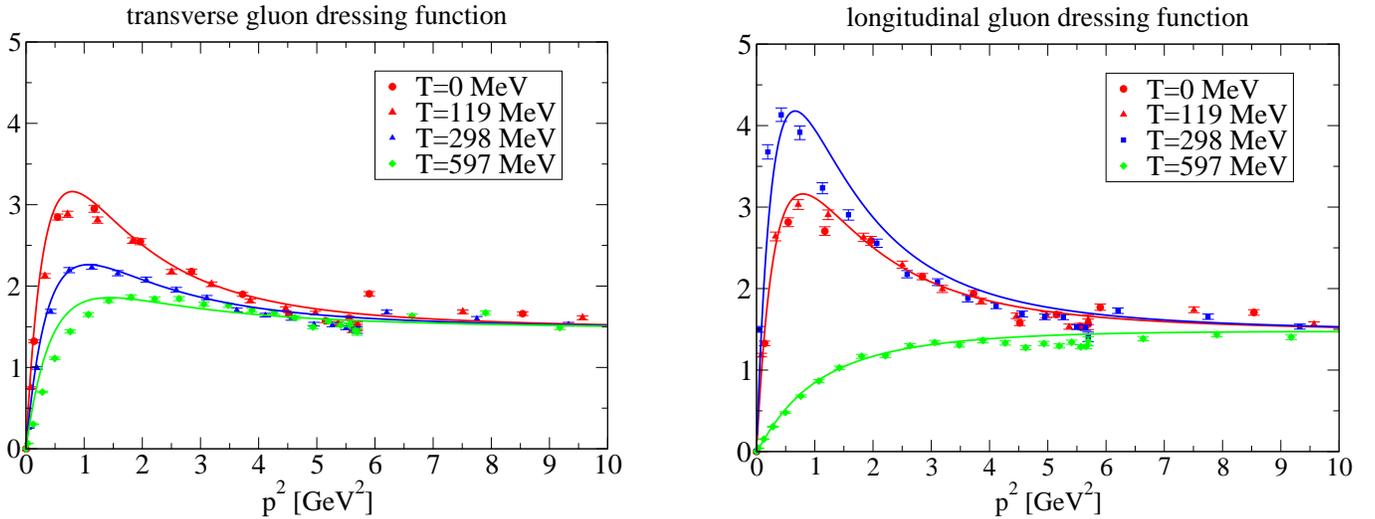

\includegraphics[width=0.95\columnwidth]{glue-trans.eps}\hfill
\includegraphics[width=0.95\columnwidth]{glue-long.eps}
\caption{Quenched $SU(2)$ lattice results for the transverse dressing
function $Z_T(q)$ and the longitudinal dressing
function $Z_L(q)$ of the gluon propagator \cite{Cucchieri:2007ta} together
with the fit functions \Eq{gluefit}. Note that the ultraviolet part
of the lattice data has not been corrected for hyper-cubic artefacts;
the fits (straight lines; see text) are adapted to the corresponding corrected 
data \cite{Sternbeck:2006cg} at temperature $T=0$.}
\label{fig:lattglue}
\end{figure*}
The Dyson-Schwinger equation for the quark propagator \Eq{quark} is
displayed diagrammatically in Fig.~\ref{fig:quarkDSE}. At
finite temperature $T$ it is given by 
\begin{widetext}
\beqa \label{DSE}
S^{-1}(\vp,\op) = Z_2 \, S^{-1}_0(\vp,\op) 
-  C_F\, \frac{Z_2 \widetilde{Z}_1}{\widetilde{Z}_3}\, g^2 T \intk \,
\gamma_{\mu}\, S(\vk,\ok) \,\Gamma_\nu(\vk,\ok,\vp,\op) \,
D_{\mu \nu}(\vp-\vk,\op-\ok) \,. \label{quark_t}
\eeqa
\end{widetext}
Here $D_{\mu \nu}$ denotes the (transverse) gluon propagator in Landau gauge and 
we have introduced a reduced quark-gluon vertex $\Gamma_\nu$, by defining
$\Gamma^{full}_{\nu,i} = i g \frac{\lambda_i}{2} \Gamma_\nu$. The 
bare quark propagator is given by 
$S^{-1}_0(p) = i \gamma \cdot p + Z_m m(\mu^2)$, where $m(\mu^2)$ is the 
renormalized current quark mass. 
The Casimir factor $C_F = (N_c^2-1)/N_c$ stems from the color trace; 
in this work we only consider the gauge group $SU(2)$.
The wave function and quark mass renormalization factors, 
$Z_2$ and $Z_m$, are determined in the renormalization 
process. The ghost renormalization factor $\widetilde{Z}_3$ is 
canceled by a corresponding factor in our model for the quark-gluon
vertex discussed below. Furthermore we used $\tilde{Z}_1=1$ for the
renormalization factor of the Landau gauge ghost-gluon vertex.
The quark dressing functions $A(\vp,\op),B(\vp,\op)$ and 
$C(\vp,\op)$ can be extracted from Eq.~(\ref{DSE}) by suitable 
projections in Dirac-space. 

In order to solve this equation we have to specify explicit expressions
for the gluon propagator and the quark-gluon vertex. At finite temperatures
the tensor structure of the Landau gauge gluon propagator contains two parts,
one transversal and one longitudinal to the heat bath. The propagator is then 
given by ($q=(\vq,\oq)$)
\begin{eqnarray}
D_{\mu\nu}(q) = \frac{Z_T(q)}{q^2} P_{\mu \nu}^T(q) 
                    +\frac{Z_L(q)}{q^2} P_{\mu \nu}^L(q) 
\end{eqnarray} 
with transverse and longitudinal projectors 
\begin{eqnarray}
P_{\mu\nu}^T(q) &=& 
   \left(\delta_{i j}-\frac{q_i q_j}{\vq^2}\right) 
   \delta_{i\mu}\delta_{j\nu}\,, \nonumber\\
P_{\mu\nu}^L(q) &=& P_{\mu \nu}(q) - P_{\mu \nu}^T(q) \,,
\end{eqnarray} 
with ($i,j=1 \dots 3$). The transverse dressing $Z_T(\vq,\oq)$ is also known as 
magnetic dressing function of the gluon, whereas the longitudinal component 
$Z_L(q)$ is called electric dressing function of the gluon propagator. At
zero temperatures Euclidean $O(4)$-invariance requires both dressing functions 
to agree, i.e. $Z_T(q)=Z_L(q)=Z(q)$.

For the momentum range
relevant for \Eq{DSE} we nowadays have very accurate solutions for the gluon
dressing function $Z(q)$ at zero temperature from both, lattice calculations 
and functional methods, see e.g. \cite{Fischer:2008uz} and references therein. 
The temperature dependence of the gluon propagator, however, is much less 
explored, see e.g. \cite{Maas:2005hs}. 
In \cite{Cucchieri:2007ta} a combined study on the lattice and 
from Dyson-Schwinger equations records a very different temperature dependence 
of the electric and magnetic parts. This is shown in Fig.~\ref{fig:lattglue}. 
The temperature effects on both the magnetic and electric dressing functions 
are such that there are almost no effects when comparing the $T=0$ result 
with $T=119$ MeV. 
Further increasing the temperature to $T=298$ MeV and $T=597$ MeV significantly
decreases the bump in the magnetic dressing function around $p^2=1$ GeV$^2$. There
is no indication that this decrease takes special notice of the critical temperature
$T_c \approx 300$ MeV for quenched QCD with gauge group $SU(2)$. The
opposite seems to be true for the electric part of the propagator. Here from 
$T=119$ MeV to $T=300$ MeV one observes a clear increase of the bump in the 
dressing function $Z_L(q)$ and a subsequent decrease when the temperature
is further raised to $T=597$ MeV. Pending further investigation it seems 
reasonable to assume that the maximum of the bump is reached at or around the 
critical temperature $T_c \approx 300$ MeV.

\begin{table}[b]
\begin{tabular}{c||c|c|c|c}
T[MeV]     & 0 & 119  & 298  & 597  \\\hline\hline
$a_{T}(T)$   & 1 & 1    & 1.34 & 1.65 \\   
$a_{L}(L)$   & 1 & 1    & 0.8  & 4.0    
\end{tabular}
\caption{Temperature dependent fit parameter of Eq.(\ref{gluefit}).\label{tab}}
\end{table}

Although the lattice data still have considerable systematic errors 
\cite{Cucchieri:2007ta} they may very well correctly represent the qualitative 
temperature dependence of the gluon propagator. We therefore use a 
temperature dependent (qualitative) fit to the data as input into the DSE; this 
fit is also displayed in Fig.~\ref{fig:lattglue} (straight lines). Note that for momenta 
around $p^2 = 6$ GeV$^2$ the lattice data show considerable systematic 
errors due to hyper-cubic artefacts. For the fits we have eliminated
these artefacts by comparison with the $T=0$ lattice result of 
Ref.~\cite{Sternbeck:2006cg}, where these errors have been corrected.
The fit functions read
\beqa \label{gluefit}
Z_{T,L}(\vq,\oq,T) &=& \frac{q^2 \Lambda^2}{(q^2+\Lambda^2)^2} \,
\left\{\left(\frac{c}{q^2+ \Lambda^2 a_{T,L}(T)}\right)^2 \right.\nonumber\\
&&\left.+\frac{q^2}{\Lambda^2}\left(\frac{\beta_0 \alpha(\mu)\ln[q^2/\Lambda^2+1]}{4\pi}\right)^\gamma\right\}\,,
\nonumber\\
\eeqa
with the temperature independent scale $\Lambda = 1.4$ GeV and 
the coefficient $c=9.8 \,\mbox{GeV}^2$. For gauge group $SU(2)$ we have
$\beta_0 = 22/3$ and $\gamma=-13/22$ in the quenched theory and we renormalize 
at $\alpha(\mu)=0.3$. The temperature dependent 
scale modification parameters $a_{T,L}(T)$ are given in table \ref{tab}.
In order to extend this fit to temperatures not given in the table we assume 
$a_{T,L}(T)$ to be temperature independent below $T=119$ MeV and only slowly 
rising above $T=597$ MeV. For $T \in [119,597]$ MeV we use cubic splines to 
interpolate smoothly between the values given in table \ref{tab}. We expect 
the systematic error of this procedure to be of the same order as the 
systematic errors inherent in the lattice data. We also inherit the scale 
determined on the lattice using the string tension 
$\sqrt{\sigma}=0.44$ GeV \cite{Cucchieri:2007ta}.  

For the quark-gluon vertex with gluon momentum $q=(\vq,\oq)$ and the quark 
momenta $p=(\vp,\op),k=(\vk,\ok)$ we employ the following temperature 
dependent model
\begin{widetext}
\beqa \label{vertexfit}
\Gamma_\nu(q,k,p) = \widetilde{Z}_{3}\left(\delta_{4 \nu} \gamma_4 
\frac{C(k)+C(p)}{2}
+  \delta_{j \nu} \gamma_j 
\frac{A(k)+A(p)}{2}
\right)\left( 							
\frac{d_1}{d_2+q^2} 			
 + \frac{q^2}{\Lambda^2+q^2}
\left(\frac{\beta_0 \alpha(\mu)\ln[q^2/\Lambda^2+1]}{4\pi}\right)^{2\delta}\right) \,,
\eeqa 
\end{widetext}
where $\delta=-9/44$ is the anomalous dimension of the vertex. Note that because of
$\gamma+2\delta=-1$ the gluon dressing function together with the quark-gluon vertex 
behaves like the running coupling at large 
momenta; this is a necessary boundary condition for any model 
interaction in the quark DSE. The dependence of the vertex on the quark dressing
functions $A$ and $C$ is motivated by the Slavnov-Taylor identity for the vertex;
it represents the first term of a generalization of the Ball-Chiu 
vertex \cite{Ball:1980ay} to finite temperatures. The remaining fit function is purely
phenomenological, see e.g. \cite{Fischer:2008wy} where an elaborate version of such 
an ansatz has been used to describe meson observables. 
Here we use $d_1 = 7.6 \,\mbox{GeV}^2$ and $d_2=0.5 \,\mbox{GeV}^2$.
A moderate variation of these parameters shifts the critical temperatures of both, 
the chiral and the deconfinement transition but leaves all qualitative 
aspects of the results presented below unchanged. 

To our knowledge the truncation scheme defined above is the first of its 
kind within the DSE-approach that implements a realistic temperature 
dependence of the gluon propagator and the quark-gluon vertex beyond 
simple ansaetze, see e.g. 
\cite{Bender:1996bm,Roberts:2000aa,Maris:2000ig,Horvatic:2007qs} 
for previous approaches. The explicit expressions of the resulting DSEs 
for the quark dressing functions including \Eq{gluefit} and \Eq{vertexfit} 
are given in appendix \ref{app:DSE}. Unless explicitly denoted otherwise 
all results presented in section \ref{sec:num} of this work are obtained 
using these DSEs. 

For the sake of comparison, however, we also employed a simplified input, 
where all temperature dependencies are solely due to the quark propagator. 
To this end we fix the temperature dependent parameter $a_{T,L}(T)$ in 
the gluon dressing function to their zero temperature values and remove
the temperature effects in the vertex using
\begin{widetext}
\beqa \label{vertexfitwoT}
\Gamma_\nu(q,k,p) = \widetilde{Z}_{3} \gamma_\nu
\left( 							
\frac{d_1}{d_2+q^2} 			
 + \frac{q^2}{\Lambda^2+q^2}
\left(\frac{\beta_0 \alpha(\mu)\ln[q^2/\Lambda^2+1]}{4\pi}\right)^{2\delta}\right) \,.
\eeqa 
\end{widetext}
with (modified) fit parameters 
$d_1 = 10 \,\mbox{GeV}^2$ and $d_2=0.2 \,\mbox{GeV}^2$ such that the chiral
transition temperature is the same as for the temperature dependent interaction. 

Finally, when investigating the effect of the details of the quark-gluon 
interaction
on the phase transition an interesting question concerns the influence of the
deep infrared behavior of the ghost and gluon propagators as well as the 
quark-gluon vertex on the phase transition. To this end we note that the
expressions (\ref{gluefit}) and (\ref{vertexfit}) for the gluon propagator 
and the quark-gluon vertex employed above correspond to decoupling type of 
infrared behavior in the sense specified in \cite{Fischer:2008uz}. The 
corresponding expressions for scaling in the infrared are given in 
appendix \ref{app:scaling}, where also corresponding numerical results for 
the temperature dependent condensates are given. As a result we find that 
the chiral and deconfinement transitions is insensitive to the question 
of scaling vs. decoupling in the deep infrared.  

\subsection{Numerical procedure \label{proc}}

{\bf Infinite volume/continuum DSEs:}\\
The numerical solutions of the DSE are found using the conventional fixed 
point iteration method. The loop integral and Matsubara sum is regularized 
using a sharp 
cutoff $\Lambda=\sqrt{5}\times10^2 \textrm{ GeV}$ such that the integration 
and summation extends to momenta and frequencies with 
$\omega^2_q+\vec{q}^2\leq\Lambda^2$. This is requisite to restore $O(4)$ 
invariance at perturbative momenta as well as in the limit of zero 
temperature. We apply a MOM renormalization scheme which defines the 
renormalization constants in such a way as to fix the full propagator 
at a certain momentum. We explicitly verified that the resulting renormalized
quark propagator is independent of the size of the ultraviolet cutoff
by increasing/decreasing the cutoff by an order of magnitude.
The renormalized quark mass $m(\mu)$ has 
contributions stemming from the explicit chiral symmetry breaking and 
from the quark condensate. Since the quark condensate is temperature 
dependent fixing the renormalized quark mass with temperature independent
renormalization conditions means that the magnitude of explicit chiral
symmetry breaking changes with temperature. In order to minimize this effect
we choose a very large renormalization point such that the
quark condensate term is irrelevant there and the temperature
effects in $m(\mu)$ can safely be ignored. The renormalization conditions
 are $C(\mu)=1$ and $B(\mu)=m(\mu)$ with $\mu=(\vec{\mu},\pi T)$ 
and $\vec{\mu}^2=10545 \textrm{ GeV}^2$. 
The choice $m(\mu)=1.85$ MeV then matches the renormalization 
conditions in Ref.~\cite{Fischer:2009wc} for mass $10 \textrm{ MeV}$ 
at $\textrm{T}=200 \textrm{ MeV}$ and $\vec{\mu}^2=20\textrm{ GeV}^2$ 
and corresponds roughly to an up quark mass.
For $|n_t|\leq 20$ the summation over Matsubara frequencies is performed 
explicitly. The remaining sum is approximated by an integral that can 
be done with standard Gauss-Legendre integration.

In the chiral limit the transition temperature for the chiral phase 
transition is unambiguously defined by the temperature at which the 
symmetry is completely restored. This is signaled by the vanishing 
of the order parameters i.e. the quark mass function at zero momentum 
or equivalently the vanishing of the chiral quark condensate. For 
explicitly broken chiral symmetry we encounter a crossover rather 
than a phase transition and the transition temperature may depend 
on its definition. It can be defined by the peak of the chiral 
susceptibility
\begin{align}
 \chi=\partial/(\partial\, m)~\langle\bar{\psi} \psi\rangle \label{def1}
\end{align}
or the renormalized chiral susceptibility
\begin{align}
 \chi_R=m^2\frac{\partial}{\partial m}
 \Big(\langle\bar{\psi}\psi\rangle_T-\langle\bar{\psi}\psi\rangle_{T=0}\Big) \label{def2}
\end{align}
which can be made dimensionless by normalizing with $T^4$ 
for details see Refs.~\cite{Karsch:2003jg,Aoki:2006br}. An 
alternative definition of the transition temperature is via 
the (normalized) slope of the condensate with temperature
\begin{align}
\tau=\sup_{T}\;\Big(-\frac{\partial~\langle\bar{\psi}\psi \rangle}{T^2\partial~ T}\Big). \label{def3}
\end{align}

The $\varphi$ dependent quark condensate is determined according to 
\begin{align} \label{trace}
 \langle\bar{\psi}\psi \rangle_{\varphi} = 
 Z_2\, N_c\, T\sum_{\omega_p(\varphi)}\int\frac{d^3p}{(2\pi)^3}\,
 \textrm{tr}_D\,S(\vec{p},\omega_p(\varphi))
\end{align}
with the conventional quark condensate obtained for $\varphi=\pi$
and multiplication with $Z_m$.
As mentioned in section \ref{dualcond}, the integral in \eq{trace}
is divergent for nonzero quark masses and needs to be regularized.
To this end we introduce a cutoff at large enough momenta where
the quark propagator in the integrand is temperature 
independent to very good approximation. The resulting regularized 
condensate is not indicative as concerns its absolute value but
retains its correct dependence on temperature $T$ and boundary 
angle $\varphi$ as long as the temperature is much smaller than 
the cutoff scale. This is the case for all results presented in the
following.

\begin{figure*}[t]
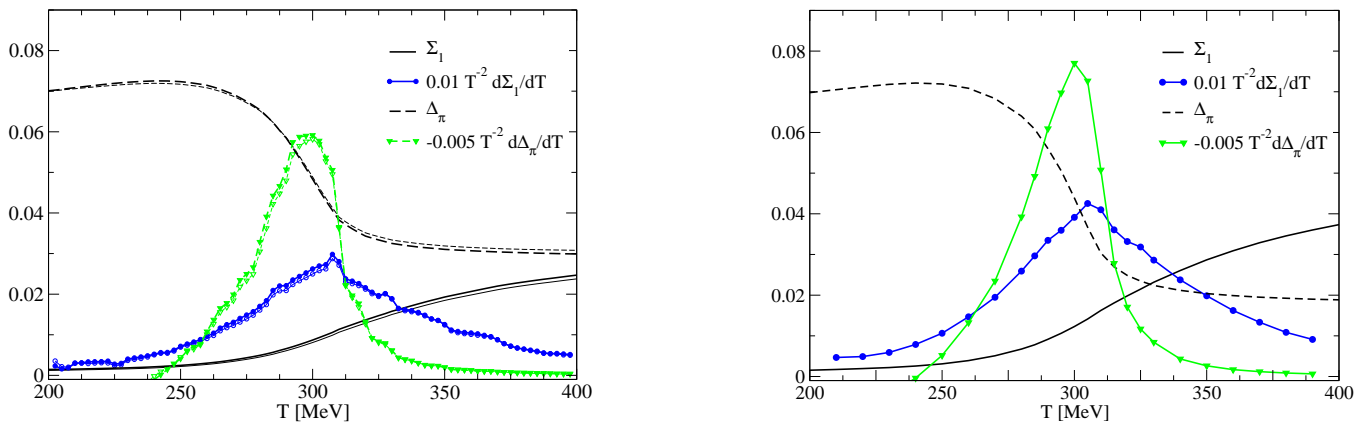

\includegraphics[width=0.9\columnwidth]{pic8.eps}\hfill
\includegraphics[width=0.9\columnwidth]{cond_dual_cond_vs_T_m10MeV.eps}
\caption{
Left diagram: Temperature dependence of the dressed Polyakov-loop 
$\Sigma_1$ and the conventional quark condensate 
$\Delta_\pi \equiv \langle \overline{\psi} \psi \rangle_{\varphi=\pi}$  
together with their derivatives for $m = 10 \,\mbox{MeV}$ evaluated
at two different ultraviolet cutoffs 
(thick lines and filled symbols for the
large cutoff; thin lines and open symbols for the smaller cutoff).
Right diagram: The same calculation in the infinite volume limit and
at an extremely large ultraviolet cutoff. Note that the ordinary condensates
in both plots have been shifted to agree at $T=200$ MeV.}
\label{res:torcont}
\end{figure*}

{\bf DSEs on a torus:}\\
In Ref.~\cite{Fischer:2009wc} first results for the ordinary and the dual chiral
condensate have been presented in a formulation of the Dyson-Schwinger equations 
on a torus.
Such a formulation has the advantage of considerable simplifications as concerns
the numerics but has the disadvantage of introducing artefacts due to the finite 
volume and lattice spacing. The calculations on a torus have been carried out at
a three-volume of $V=(5 \,\mbox{fm})^3$. This is sufficiently large to avoid
significant volume effects even for quark masses as small as 
$m(\vec{\mu}^2=20 \,\mbox{GeV}^2) = 10 \,\mbox{MeV}$ used in Ref.~\cite{Fischer:2009wc}.
For even larger volumes one obtains small changes of the order of a few percent
in the chiral condensate below the critical temperature and much smaller effects
for larger temperatures; the value of the chiral transition temperature is
unaffected. In addition there are effects due to the finite lattice spacing
in temporal and spatial directions and the fact that the renormalized 
quark mass $m=B(\pi T,\vec{\mu})/C(\pi T,\vec{\mu})$ has been treated as a 
constant with respect to variations in temperature \cite{Fischer:2009wc}. 
This is certainly correct for a very large renormalization point $\vec{\mu}$, 
but introduces temperature dependent 
artefacts for the value $\vec{\mu}^2 = 20$ GeV used in the torus 
calculation. In this work we remove these artefacts by the following
procedure: we first solve the Dyson-Schwinger equations in the infinite volume/continuum
limit, i.e. in the formulation introduced in section \ref{sec:DSE}. These equations
are renormalized at a very large renormalization point and consequently
temperature effects in the renormalization procedure are absent. From these
results we read off the corresponding, temperature dependent renormalization 
conditions at the small renormalization point $\vec{\mu}^2=20 \,\mbox{GeV}^2$ 
of our torus formulation. These are then applied to the calculations on the torus. 
This procedure is justified if and only if the truncation of the quark-DSE respects
multiplicative renormalizability, which is the case for our truncation scheme.
As we will see below, this change in the renormalization procedure leads
to a significant change in the deconfinement transition temperature on the torus
as compared to the one extracted in Ref.~\cite{Fischer:2009wc}.

The numerical details for the calculation on a torus are similar to the
ones in the zero temperature limit, discussed at length in 
Refs.~\cite{Fischer:2002eq}. As concerns corrections for hyper-cubic 
artefacts we employ the 'cutting-the edges' prescription developed
in \cite{Fischer:2002eq} in the three spatial directions of the torus.
For the temporal direction we take into account as many Matsubara modes
as necessary to arrive at approximately the same cutoff scale as for the
spatial directions. This results e.g. in $n_t=8$ temporal Matsubara modes
at $T=200$ MeV. 

These prescriptions are adapted to minimize volume and discretization artefacts 
when comparing solutions from DSEs on a torus with the corresponding results 
in the infinite volume/continuum limit. 
At this point we do not claim that the remaining
volume and spacing effects are similar to the corresponding ones in lattice
calculations. Nevertheless it is interesting to investigate the details of
the volume and cutoff-dependence of our torus results systematically. While
first results of this comparison are presented in the next section, a more
detailed analysis is deferred to future work.

\section{Numerical results \label{sec:num}}

\subsection{The infinite volume and continuum limit}

\begin{figure*}[t]
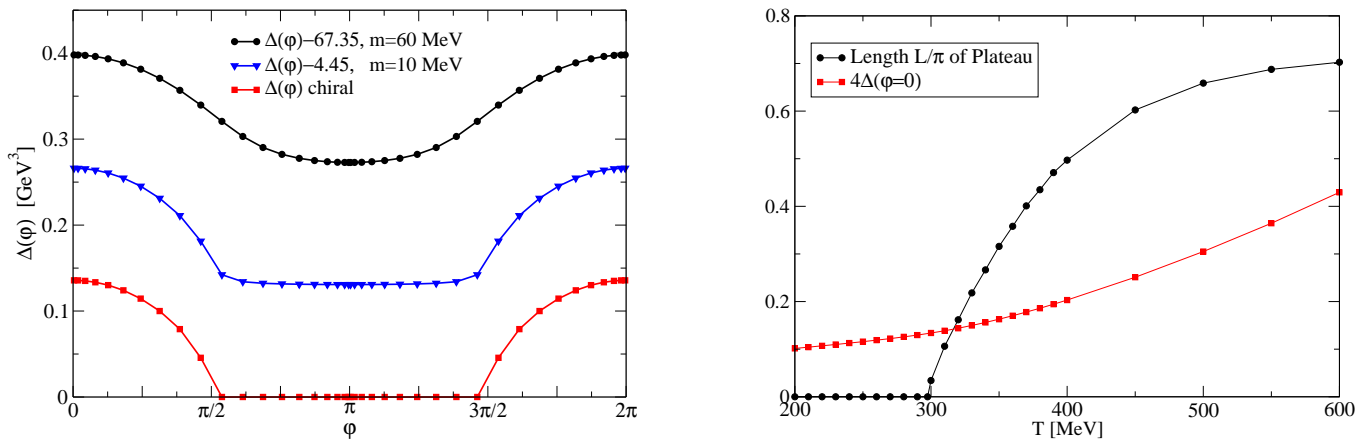

\includegraphics[width=0.97\columnwidth]{Dual_Kond_vs_phi.eps}\hfill
\includegraphics[width=0.91\columnwidth]{T_Rand_Plateau.eps}
\caption{Left diagram: Angular dependence of the quark condensate 
evaluated at two different quark masses and in the chiral limit
at $T=400$ MeV.
Right diagram: Temperature dependence of the half-width $L$ of the
plateau in the chiral quark condensate in units of $\pi$ (i.e.
$L/\pi = 1$ means that the plateau extends over all angles 
$\varphi=0 \dots 2\pi$) and temperature dependence of the chiral 
quark condensate $\Delta(\varphi)$ at periodic
boundary conditions $\varphi=0$.
}
\label{res:chiral}
\end{figure*}

We first compare results from DSEs on a torus with results from 
the infinite volume/continuum DSEs at a quark mass that roughly 
corresponds to an up-quark. The resulting temperature dependent 
ordinary and dual condensates as well as their (normalized) temperature 
derivatives are shown in Fig.~\ref{res:torcont}. The left hand diagram 
displays the torus results at two different ultraviolet cutoffs 
$p_{max}=5.96,7.45$ GeV corresponding to lattice spacings of 
$a=\pi/p_{max}\approx 0.1,0.08$ fm, respectively. The right hand diagram 
shows our results in the continuum. Compared to Ref.~\cite{Fischer:2009wc} 
we use a modified mass renormalization scheme on the torus as described 
in the last subsection. Furthermore we normalize the temperature
derivatives with $1/T^2$ to obtain dimensionless quantities. The last 
change affects both transition temperatures only by 1-2 MeV. The effects 
of the modified mass renormalization procedure can be assessed by comparing 
our results with the ones shown in Fig.~3 of Ref.~\cite{Fischer:2009wc}. 
We obtain a more pronounced chiral transition with a narrower peak in the 
temperature derivative of the ordinary chiral condensate. The corresponding 
transition temperature $\tau \approx 300$ MeV is unchanged. As for the dual 
condensate, or dressed Polyakov loop, the peak of the temperature derivative 
is shifted to smaller temperatures, i.e. from $T_{dec} \approx 320$ MeV
in Ref.~\cite{Fischer:2009wc} to our value $T_{dec} \approx 308$ MeV for
both cutoffs. Despite this shift, the deconfinement transition remains 
separate from the chiral transition.

This finding is also visible in the infinite volume/continuum limit. Although 
there are still differences in the details of the temperature dependence of 
the ordinary and dual condensate between torus and continuum DSEs, the related 
peaks in the temperature derivatives are similar; here we find 
$\tau \approx 301$ MeV and $T_{dec} \approx 308$ MeV, i.e. still (slightly) 
different transition temperatures. This gap reduces further when the
corresponding transition temperatures from the chiral susceptibilities 
$\chi_R/T^4$ and $\chi_R$ are considered, see table \ref{crit_temp}.
\begin{table}[b]
\begin{tabular}{c|c|c|c}
 $\tau$  & $T_{\chi_R/T^4}$ & $T_{\chi_R}$ & $T_{dec}$   \\\hline\hline
$301(2)$ & $304(1)$         & $305(1)$     & $308(2)$
\end{tabular}
\caption{Transition temperatures in the infinite volume/continuum limit 
for the different definitions Eqs.~(\ref{def1}), (\ref{def2}) and (\ref{def3}).
These values have been extracted from numerical results evaluated using
temperature steps of one MeV (not shown in Fig.\ref{res:torcont}).\label{crit_temp}}
\end{table}

In general, our results in the infinite volume/continuum limit agree qualitatively 
with the ones on a torus and almost quantitatively as concerns the transition 
temperatures. The most prominent remaining discrepancy is seen in the relative 
decrease of the ordinary condensate and the relative increase of the dual condensate 
below and above the transition temperature. This difference is a cutoff effect, as 
can be inferred from comparing results for the two different lattice spacings in 
the left plot of Fig.~\ref{res:torcont}. In particular we found that cutoff effects
in the quark-DSE only play a very minor role, whereas the cutoff effects in the
condensate equation (\ref{trace}) are most significant. These effects will be 
explored further in future work.

In general we wish to emphasize that the present calculation, although carried
out with quenched lattice results for the gluon propagator, is in itself
not strictly quenched: our ansatz for the quark-gluon vertex is too simple
to strictly represent the quenched theory. This can be seen from the fact that
the dressed Polyakov-loop is not strictly zero below the deconfinement
transition. Consequently we do not observe the second order deconfinement
phase transition expected from quenched $SU(2)$ Yang-Mills theory but a, 
more or less rapid, crossover at $T_{dec} \approx 308$ MeV\footnote{Note that
even if our vertex were strictly quenched it is not clear whether the
lattice input for the gluon propagator is precise enough to allow for an
observation of the second order phase transition expected in 
$SU(2)$ Yang-Mills theory}. On the other hand,
as concerns the chiral properties of our model, a prominent feature of the
quenched theory not reproduced by our framework is the appearance of 
quenched chiral logarithms in the chiral condensate. These are well-known
to be generated by $\eta'$ hairpin diagrams, which are not represented by
our vertex ansatz. For the present investigation this is
more an advantage than a drawback. Quenched chiral logarithms are most 
notable in the chiral limit, where they lead to a singularity in the chiral 
condensate. Since we do not encounter this singularity we are in a position
to investigate both, the ordinary and the dual condensate in the chiral limit.
\begin{figure*}[t]
\includegraphics[width=0.9\columnwidth]{Kond_Dual-Kond_vs_Temp_chiral.eps}\hfill
\includegraphics[width=1.1\columnwidth]{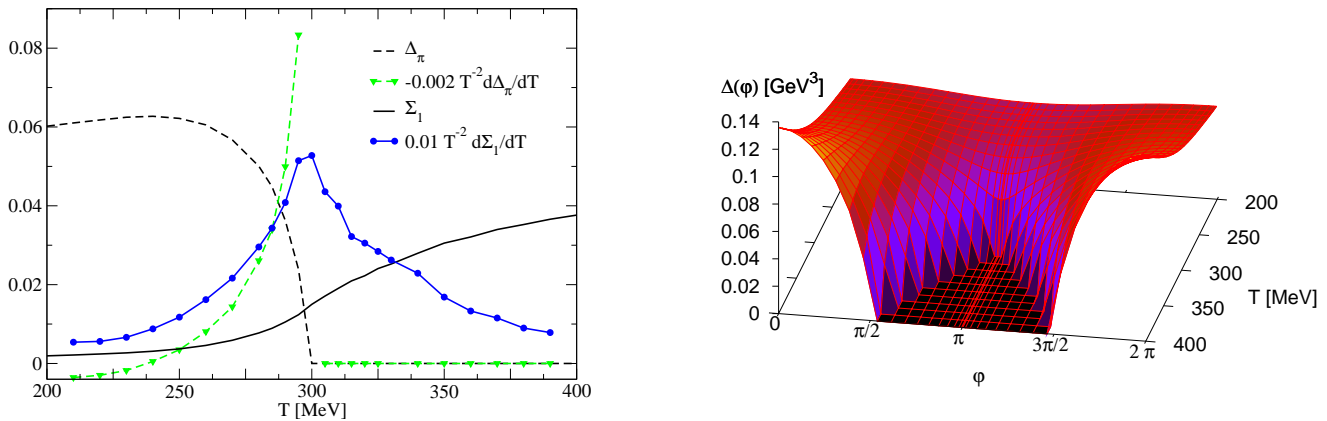}
\caption{Left diagram: Temperature dependence of the conventional 
and dual condensates as well as their temperature derivatives in the chiral limit.
Right diagram: A 3d-plot of the angular and temperature
dependence of the chiral quark condensate.
}
\label{res:chiral2}
\end{figure*}

\subsection{The chiral limit\label{sec:chiral}}

In Fig.~\ref{res:chiral} we study the dependence of the quark
condensate on the boundary
angle $\varphi$. In the left diagram we compare the angular dependence
of the condensate at $T=400$ MeV for two different quark masses and in the chiral limit.
As already noted in Ref.~\cite{Fischer:2009wc} we clearly see a broadening
in the central dip of the graphs with decreasing quark mass. This can be
readily understood from the loop expansion of the quark condensate,
Eq.~(\ref{loop}), which is applicable at finite quark masses and under
presence of an ultraviolet regulator for the condensate, cf. section \ref{proc}. 
At sufficiently large quark masses large loops are suppressed  
by powers of $1/m$. As a result only loops winding once around the torus
should contribute in \Eq{loop} and the resulting angular behavior of the
condensate should be proportional to $\cos(\varphi)$. Indeed, this is what we see:
the result for our largest quark mass can well be fitted by only few terms in an expansion  
$\Delta(\varphi) = \sum_{n=0}^N a_n \cos(n\varphi)$ and the first term
is by far the largest contribution. For smaller quark masses we observe
also sizeable contributions from terms $\cos(n\varphi)$ with $n>1$. In the plot,
these contributions are responsible for the flat area around the antiperiodic
boundary angle $\varphi=\pi$. Approaching the chiral limit this area becomes
flatter and finally develops a derivative discontinuity
at two finite values of $\varphi = \pi \pm L$. These indicate the breakdown of 
the loop expansion \Eq{loop} in the chiral limit. Note that this
is also the limit where the condensate is free of quadratic ultraviolet
divergencies and the continuum limit can be taken; another reason why the
loop expansion \Eq{loop} is no longer applicable.

In the plot on the right hand side of Fig.~\ref{res:chiral} we show the temperature
dependence of the half-width $L$ of the plateau in the chiral quark condensate 
in units of $\pi$. The width of the plateau is zero below the critical 
temperature $T_c$ and rises quickly above, spreading up to 
$L/\pi=0.6$ for temperatures in the region $T \approx 2 T_c$. This monotonic 
rise then slows down considerably and seems to converge to a finite value 
smaller than one. However, from the presently available results up to 
$T \approx 2.5 T_c$ one cannot completely exclude that the plateau finally 
extends over the whole range of boundary angles $\varphi=0 \dots 2\pi$ for 
$T \rightarrow \infty$. 

The behavior of the chiral quark condensate at periodic boundary conditions,
$\Delta(\varphi=0)$, is also a monotonic function of temperature as shown in the
same plot. From a naive dimensional analysis of the condensate one may expect
that it rises proportional to the third power of the temperature. This is
however not correct as explicitly shown in our analytical analysis
in appendix \ref{app:largeT}. There we demonstrate that  
\beq
\Delta_{\varphi=0}(T) \sim T^2  \hspace{1cm}\text{for} \hspace{2mm}T\gg T_c \,.
\eeq
To extract this behavior also from our numerical data we fitted a 
polynomial of degree three in $T$: the result, 
$\Delta_{\varphi=0}(T \gg T_c) = 10^{-5} + 0.045(1) T + 0.73(1) T^2 + 10^{-7} T^3$,
for temperatures $T > 2 T_c$ clearly agrees with the analytical analysis.
This scaling behavior also agrees with corresponding results of Ref.~\cite{heid}.
In addition it may be interesting to note that this $T^2$-scaling of the
condensate at periodic boundary conditions agrees with the naive scaling
of the chiral condensate of the resulting three-dimensional theory in the
infinite temperature limit.

In the diagram of the left hand side of Fig.~\ref{res:chiral2}
we plot the temperature dependence of the ordinary and the chiral condensate
as well as their temperature derivatives in the chiral limit. One clearly 
observes a second order chiral phase transition. The temperature derivative 
of the ordinary condensate diverges at the corresponding critical temperature 
$T_c = \tau = T_{\chi_R/T^4} = 298(1)$ MeV. 
Above $T_c$ the ordinary condensate is strictly zero. For the
deconfinement transition we observe a qualitatively similar behavior as for
the case with finite quark mass discussed in the last subsection. The
corresponding transition temperature, however, has moved to a lower value, 
$T_{dec} = 299(3)$ MeV, and coincides almost precisely with the chiral critical
temperature. This is in agreement with corresponding extrapolations within
lattice QCD \cite{Karsch:1998ua}. We regard this as one of the main results
of the present investigation.

\begin{figure*}[t]
\includegraphics[width=1.1\columnwidth]{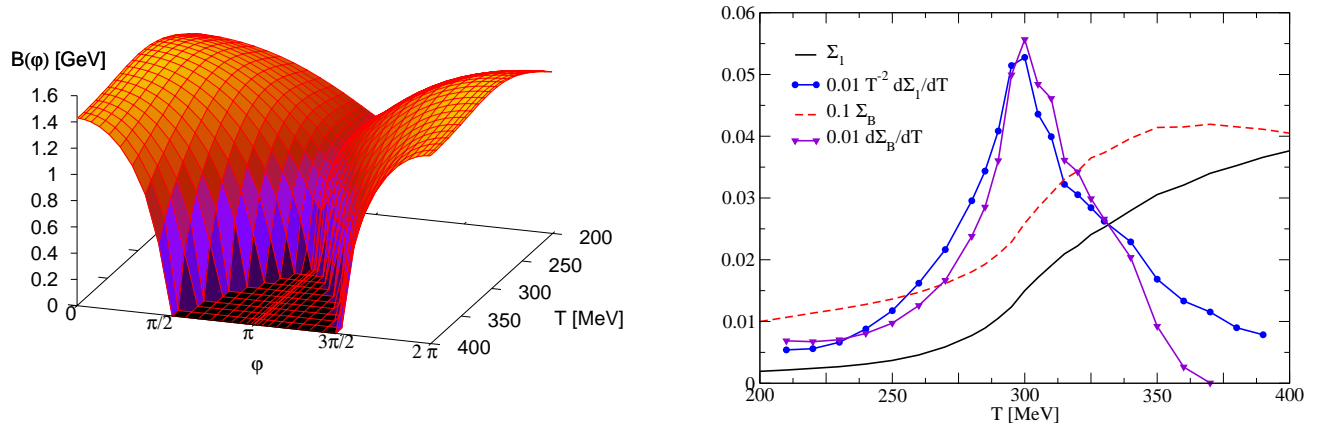}\hfill
\includegraphics[width=0.9\columnwidth]{dual_cond_dual_b_vs_T_chiral.eps}
\caption{
Left diagram: A 3d-plot of the angular and temperature
dependence of the quark scalar dressing function evaluated at zero momentum
and lowest Matsubara frequency. Note that for $T> 400$ MeV 
(not shown in the plot) the scalar
quark dressing function at periodic boundary conditions starts to rise
linearly with temperature, $B_{\varphi=0, p=0}(T) \sim T$. 
Right diagram: Comparison of the temperature dependence of the 
corresponding dual scalar quark dressing and the dual condensate.
}
\label{res:scalar}
\end{figure*}
A combined plot of the angular and temperature dependence of the chiral
limit condensate is shown in the right hand diagram of Fig.~\ref{res:chiral2}.
We clearly observe a different evolution of the quark condensate at
different boundary angles $\varphi$. Whereas at periodic boundary
conditions $\varphi=0,2\pi$ the condensate is monotonically rising, cf. 
the curve in the right diagram in Fig.~\ref{res:chiral}, it clearly shows 
signals for a phase transition for boundary angles close to antiperiodic 
boundary conditions $\varphi=\pi$, cf. the corresponding curve in the 
left diagram of Fig.~\ref{res:chiral2}. Indeed for angles around $\varphi=\pi$
one could (formally) define a $\varphi$-dependent transition temperature
that is related to the width $L$ of the plateau. 

Finally we would like to make two comments. 
First, the corresponding dependence of the
massive quark condensate as function of $(T,\varphi)$ is similar to that of its 
chiral counterpart. The most pronounced difference is that there is no 
plateau around $\varphi=\pi$, but instead a cosine-shaped dip as can be 
inferred from the left diagram in Fig.~\ref{res:chiral}.  
Second, we also determined the chiral condensate using the simplified, 
temperature independent quark-gluon interaction, \Eq{vertexfitwoT}, and 
a gluon propagator fixed at $T=0$. This form of the interaction allows 
for temperature effects in the quark propagator only, while retaining the
chiral critical temperature $\tau \approx 300$ MeV.
The corresponding result is qualitatively similar to the one with full
temperature dependence in the interaction. However, both transitions
are much less pronounced. The chiral condensate starts to decrease and 
the dual condensate starts to increase at much lower temperatures.
Correspondingly we obtain broad curves for the temperature derivatives.
We conclude that a quantitative study of the details of the phase 
transitions of QCD has to take into account the temperature effects in the 
interaction.

\subsection{The dual scalar quark dressing function\label{sec:resdual}}

We now come back to the dual scalar quark dressing introduced by \Eq{dualskalar}.
Our numerical results for the angular and temperature dependence 
of the scalar quark dressing function $B(0)$ is shown in the left diagram 
of Fig.~\ref{res:scalar}. The resulting pattern is similar to the one for 
the quark condensate, cf. Fig.~\ref{res:chiral2}, in particular as concerns 
the opening of a zero plateau around antiperiodic boundary conditions above
the critical temperature. For temperatures below $T_c$ we observe a more 
distinct $\varphi$ dependence as for the dual condensate, again indicating the non-quenched nature of our 
model for the quark-gluon vertex as discussed above. Above $T_c$ the scalar
quark dressing function $B_{\varphi=0}(0)$ at periodic boundary conditions stays
constant for a while but then starts to rise for temperatures above $2 T_c$
(not shown in the 3d-plot). Our fit to the data leaves us with 
\beq
B_{\varphi=0, p=0}(T) \sim \sqrt{T} \hspace{1cm}\text{for} \hspace{2mm}T\gg T_c
\eeq
in agreement with our analytical
analysis in appendix \ref{app:largeT} and corresponding results of 
Ref.~\cite{heid}. 

The data for the dual scalar quark dressing are shown on the right
hand diagram of Fig.~\ref{res:scalar} and compared to the dual condensate.
Again, the temperature dependence of both quantities is very similar. The temperature
derivatives clearly peak at precisely the same transition
temperature, indicating the deconfinement transition. Also, the rate of change
at and around the transition temperature is quantitatively similar, although the
signal is slightly stronger for the dual scalar dressing function $\Sigma_B$.
From a systematic perspective the dual condensate, or dressed Polyakov loop, has 
the advantage of a direct connection to the ordinary Polyakov loop in the
limit of static quarks. However, the dual scalar quark dressing function has
the advantage of being a well defined quantity in the continuum limit. After 
renormalization of the Dyson-Schwinger equation of the quark propagator this 
quantity stays finite when the cutoff is sent to infinity also for finite quark 
masses. Both quantities are promising candidates for a further study of the 
deconfinement transitio of QCD.

\section{Summary and Conclusions \label{sec:sum}}

In this work we have investigated the chiral and the deconfinement
transition of quenched QCD. From the temperature dependent quark 
propagator we extracted the order parameter for the chiral transition, 
the quark condensate, and order parameters for the deconfinement 
transition, the dressed Polyakov loop and the dual scalar quark dressing.
We determined the quark propagator from its Landau gauge
Dyson-Schwinger equation in the imaginary time Matsubara formalism.
We emphasize once more that our truncation scheme, introduced in 
Ref.~\cite{Fischer:2009wc}, implements a realistic temperature dependence 
of the gluon propagator and takes into account temperature effects in  
the quark-gluon vertex. It therefore goes beyond the simple models 
used in previous works.

When comparing results from the infinite volume/continuum DSEs with
corresponding ones on a torus we found interesting effects due to
different renormalization prescriptions for the quark propagator:
the deconfinement transition temperature on a torus is shifted to 
smaller temperatures, i.e. from $T_{dec} \approx 320$ MeV to 
$T_{dec} \approx 308$ MeV when artificial temperature effects in 
the renormalized quark mass are avoided. This value agrees precisely 
with $T_{dec} \approx 308(2)$ MeV in the infinite volume/continuum 
limit reflecting the fact that the remaining cutoff artefacts on 
our torus do not affect the transition temperature. The same is true
for the chiral transition temperature. We find $\tau = 301(2)$ MeV 
from the temperature derivative of the chiral condensate and 
$T_{\chi_R/T^4} = 304(1)$ MeV from the normalized chiral susceptibility. 
In summary we conclude that the chiral and deconfinement transition 
temperatures are only slightly different for finite quark masses. 
This finding will be further tested in future work using more precise 
data for the gluon propagator. 

In the chiral limit this agreement becomes exact (within numerical errors). 
We find a second order chiral phase transition at $T_c = 298(1)$ MeV 
and a similar temperature for the deconfinement transition, 
$T_{dec} =299(3)$ MeV. It is worth to emphasize again that both
transition temperatures are extracted from the properties of the
quark propagator and therefore related to the underlying properties of the
Dirac operator. Therefore in a sense discussed in more detail in
Refs.~\cite{Gattringer:2006ci,Bruckmann:2006kx,Synatschke:2007bz,Synatschke:2008yt,heid}
the chiral and the deconfinement transition are closely connected.

The framework used in this work is quenched $SU(2)$ Yang-Mills theory.
Our transition temperatures may be translated into the corresponding ones
of quenched $SU(3)$ QCD using the relations $T_c/\sqrt{\sigma}=0.709$
($SU(2)$) and $T_c/\sqrt{\sigma}=0.646$ ($SU(3)$) between the respective 
critical temperatures and the string tension \cite{Fingberg:1992ju}. 
The resulting transition temperatures are then 
$T_{\chi_R/T^4} \approx 277$ MeV and $T_{dec} \approx 281$ MeV 
for the massive case and 
$T_{\chi_R/T^4} \approx T_{dec} \approx 272$ MeV in the chiral limit.
In order to work in the full, unquenched theory we would have to take
into account quark-loop effects in the gluon propagator and meson effects
in the quark-gluon vertex \cite{Fischer:2008wy}. These effects will shift
the transitions temperatures below $T=200$ MeV, see \cite{Bazavov:2009zn,Aoki:2009sc}
for latest results for $N_f=2+1$ quark flavors. As concerns the dual 
condensate and scalar dressing function in the unquenched formulation 
one needs to carefully take into account effects due to the 
Roberge-Weiss symmetry \cite{Roberge:1986mm}. This is because of the formal 
similarity of the continuous boundary conditions for the quark field to an 
imaginary chemical potential, see \cite{heid} for details. An investigation 
of these effects as well as a detailed study of volume and discretization 
artefacts is deferred to future work.

{\bf Acknowledgments}\\
We thank Falk Bruckmann, Christof Gattringer, Erwin Laermann,
Jan Pawlowski, Rob Pisarski, Lorenz von Smekal and Wolfgang Soeldner 
for discussions. We are grateful to Axel Maas 
for discussions and for making the lattice data of 
Ref.~\cite{Cucchieri:2007ta} available. This work has 
been supported by the Helmholtz Young Investigator 
Grant VH-NG-332 and by the Helmholtz Alliance HA216-TUD/EMMI.

\vspace*{1cm}
\begin{appendix}
\section{Explicit form of the DSEs for the quark dressing functions\label{app:DSE}}

\noindent
{\bf DSEs in the infinite volume/continuum limit:}\\
\noindent
The explicit form of the DSEs in the infinite volume/continuum
limit in the truncation specified in section \ref{trunc} is given by
\begin{widetext}
\beqa
A(k)&=&Z_2+\frac{Z_{2}\,g^2}{\vec{k}^2}C_F\,
T\sum_{n_q}\int\,\frac{d^3q}{(2\pi)^3}\,
\frac{1}{(\omega_{q}^2\,C^2(q)+\vec{q}^2 A^2(q)+B^2(q))}\times \nonumber\\
&& \Bigg\lbrace C(q)\,\Delta^L(p)\;\omega_{q} (\omega_{q}-\omega_{k})\;
\frac{\vec{q}^2-\vec{p}^2-\vec{k}^2}{p^2}\;\frac{\mC(q,k)+\mA(q,k)}{2} \nonumber\\
&&+\frac{A(q)}{2}\bigg[\Delta^L(p)\,\frac{\vec{k}^2+\vec{q}^2-\vec{p}^2}{p^2}\,
\Big(\mC(q,k)\,\vec{p}^2+\mA(q,k)\,\omega_{p}^2\Big)
+\Delta^L(p)\,\frac{(\vec{q}^2-\vec{k}^2)^2-\vec{p}^4}{p^2}\,\mA(q,k)   \nonumber\\
&&+\Big(\Delta^T(p)-\Delta^L(p)\Big)\,
\frac{(\vec{q}^2-\vec{k}^2)^2-\vec{p}^4}{\vec{p}^2}\,\mA(q,k)\bigg]\Bigg\rbrace~,\label{A}
\eeqa
\beqa
 B(k) &=& Z_2\,Z_m\,m+Z_{2}\,g^2\,C_F\,T\sum_{n_q}\int\,\frac{d^3q}{(2\pi)^3}\;
 \frac{B(q)}{\omega_{q}^2\,C^2(q)+\vec{q}^2\,A^2(q)+B^2(q)}\times\nonumber\\
&&\Bigg(\Delta^L(p)\frac{\omega_{p}^2\,\mA(q,k)+\vec{p}^2\,\mC(q,k)}
{p^2}+2\,\Delta^T(p)\,\mA(q,k)\Bigg)~,\label{B}\label{Beq}
\eeqa
\beqa
 C(k)&=&Z_2+\frac{Z_{2}\,g^2}{\omega_{k}}\,C_F\,T\sum_{n_q}\int\,\frac{d^3q}{(2\pi)^3}\;
 \frac{1}{\omega_{q}^2\,C^2(q)+\vec{q}^2\,A^2(q)+B^2(q)}\times\nonumber\\
&&\Bigg(\omega_{q}\,C(q)\bigg[2\,\Delta^T(p)\,\mA(q,k)+\Delta^L(p)\,\Big(\mA(q,k)\,
\frac{\omega_{p}^2}{p^2}-\mC(q,k)\,\frac{\vec{p}^2}{p^2}\Big)\bigg]\nonumber\\
&&+\omega_{p}\,A(q)\,\Delta^L(p)\,\frac{\vec{p}^2+\vec{q}^2-\vec{k}^2}{2p^2}\,
\Big(\mC(q,k)+\mA(q,k)\Big)\Bigg)\label{C}
\eeqa
with $q = (\vq,\oq)$, $k = (\vk,\ok)$, $p = (\vp,\op)=q-k$, and
the gluon propagator
\beq
\Delta^{T}(p)=\frac{Z_T(p)}{p^2},\qquad \Delta^{L}(p)=\frac{Z_L(p)}{p^2}
\eeq
and the vertex dressing functions
\beq
\mC(q,k)= \Gamma(q-k)\,\frac{C(q)+C(k)}{2}  \hspace*{1cm}\text{and}\hspace*{1cm} 
\mA(q,k)= \Gamma(q-k)\,\frac{A(q)+A(k)}{2}
\eeq
with
\beq
\Gamma(q) = \left(							
\frac{d_1}{d_2+q^2} 			
 + \frac{q^2}{\Lambda^2+q^2}
\left(\frac{\beta_0 \alpha(\mu)\ln[q^2/\Lambda^2+1]}{4\pi}\right)^{2\delta}\right)\,.
\eeq
\end{widetext}

\noindent
{\bf DSEs on a torus:}\\
\noindent
The corresponding form of the DSEs on a torus is similar to the one given in 
\Eq{A},\Eq{B} and \Eq{C} with the exception that the three-dimensional
momentum integral is replaced by a sum over momenta. In the three Cartesian 
spatial directions we thus have Matsubara sums counting momenta 
$\vp_{\bf n} = \sum_{i=1..3} (2\pi/L)(n_i) \hat{e}_i$, where $\hat{e}_i$ 
are Cartesian unit vectors in Euclidean momentum space. In \cite{Fischer:2002eq}
a technique of rearranging the three Cartesian sums into a sum over
hyper-spheres and a second sum over the individual momenta on each
hyper-sphere has been developed that is convenient for the numerical treatment 
of the DSEs. This sum is given by
\beq
\frac{1}{L^3} \sum_{n_1,n_2,n_3} (\cdots) \: =  \frac{1}{L^3} \sum_{j,m} \:(\cdots) \,, 
\eeq 
where $j$ counts spheres with $\vec{p}^2=\textrm{const}$, and $m$ 
numbers the grid points on a given sphere. The corresponding momentum 
vectors are denoted by $\vp_{j,m}$. 

\section{Large temperature scaling of $B_{\varphi=0}$ and 
$\langle \overline{\psi} \psi \rangle_{\varphi=0}$  \label{app:largeT}}

In this section we analyze the large temperature scaling of the
scalar quark dressing function $B_{\varphi=0}(p,T)$ and the quark condensate
$\langle \overline{\psi} \psi \rangle_{\varphi=0}(T)$ with temperature. 
To this end we first analyze the large temperature limit of the DSE \Eq{Beq}
for the scalar quark dressing function. With periodic boundary conditions
$\varphi=0$ only the zeroth Matsubara frequency $\oq=0$ contributes in the Matsubara
sum; all others are suppressed by powers of the temperature $T$. Furthermore
at large enough temperatures the temperature effects in the gluon dressing
functions and the vector dressing functions $A$ and $C$ of the quark propagator
and quark-gluon vertex are arbitrary weak and therefore can be neglected.
Working in the chiral limit we then have  
\beqa
 B(\vk,T) &\sim& T \int\,\frac{d^3q}{(2\pi)^3}\;
 \frac{B(\vq,T)}{\vec{q}^2+B^2(\vq,T)}\frac{3\Gamma(\vq-\vk)}{(\vq-\vk)^2}\,\nonumber\\
 &\sim& T \int\,dq \frac{B(\vq,T)}{\vec{q}^2+B^2(\vq,T)}\,. \label{eq1}
\eeqa
The second line follows from the first line because of the temperature
independence of the terms in the angular integral. The resulting integral
in \Eq{eq1} is dominated from low momenta $\vec{q}^2 < B^2(\vq,T)$ and 
consequently the integrand scales like $1/B(\vq,T)$ leading to
\beq
B(\vk,T) \sim \sqrt{T} \hspace{1cm}\text{for} \hspace{2mm}T\gg T_c \,,
\eeq 
in agreement with our numerical results discussed in section \ref{sec:resdual} and
corresponding results in Ref.~\cite{heid}.
We also confirmed that indeed the scaling of $B(\vk,T)$ with $T$ is independent
of the three-momentum $\vk$.
\begin{figure*}[t]
\includegraphics[width=0.9\columnwidth]{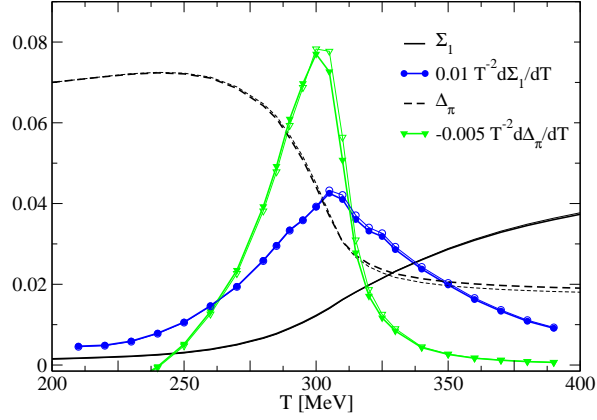}
\caption{Temperature dependence of the dressed Polyakov-loop 
$\Sigma_1$ and the conventional quark condensate 
$\Delta_\pi \equiv \langle \overline{\psi} \psi \rangle_{\varphi=\pi}$  
together with their derivatives for $m = 10 \,\mbox{MeV}$ in the infinite 
volume/continuum limit. The thick lines, already shown in the right diagram of Fig.~\ref{res:torcont}, correspond to the decoupling type of interaction
whereas the thin lines are the scaling type of interaction Eqs.~(\ref{mod0}), (\ref{mod}) and (\ref{mod2}).}
\label{scalvsdec}
\end{figure*}

Having identified the large temperature scaling of $B(\vk,T)$ we now analyze the
integral for the quark condensate \Eq{trace} given by 
\beqa 
 \langle\bar{\psi}\psi \rangle_{\varphi=0} &=& 
 4\,Z_2\, N_c\, T\sum_{\omega_q}\int\frac{d^3q}{(2\pi)^3}\,
 \frac{B}{\oq^2 C^2 
 + \vec{q}^2 A^2+B^2}\,,\nonumber\\
 &\sim& T \int dq \vq^2 \frac{B(\vq,T)}{\vec{q}^2 + B^2(\vq,T)} \label{eq2}
\eeqa
Here the second line is obtained by again neglecting the temperature scaling 
of the vector dressing functions $A,C$ and the contributions of all but the 
lowest Matsubara frequency. The resulting integral has an extra factor
of $\vq^2$ as compared to the one in \Eq{eq1}. It is therefore not dominated
by momenta smaller than $B^2$. Recalling that $B(0,T) \sim B(\vq,T) \sim \sqrt{T}$
we proceed by the integral transformation
$q \rightarrow q'B(0,T)$ which leads to
\beqa
 \langle\bar{\psi}\psi \rangle_{\varphi=0} 
 &\sim&T \int dq' \vec{q'}^2 B^3(0,T) \times \nonumber\\ 
 &&\frac{B(\vq,T)}{B^2(\vq,T)\left(\vec{q'}^2 B^2(0,T)/B^2(\vq,T) + 1\right)}\,, \nonumber\\
 &\sim& T^2\,, \hspace{1cm}\text{for} \hspace{2mm}T\gg T_c\,. 
\eeqa
This result is again in excellent agreement with our numerical results 
discussed in section
\ref{sec:chiral}.

\section{Scaling vs. decoupling \label{app:scaling}}

In this appendix we demonstrate that the deep infrared behavior of the gluon
propagator and the quark-gluon vertex is not related to the details of
the deconfinement transition at $T_{dec}$. The reason why we are interested 
in this question is an ongoing discussion over the past years on the status 
of two possible types of analytical infrared solutions of Yang-Mills theory.
These two types of solutions are characterized by a one parameter family of boundary
conditions on the dressing function of the ghost propagator, $G(p^2)$, at
zero momentum. If $G(p^2=0) = \infty$ one observes the 'scaling' solution of
infrared Yang-Mills theory, characterized by power laws in all one-particle
irreducible Green's functions of the theory 
\cite{Alkofer:2004it,Fischer:2006vf,Alkofer:2008jy}. The other possibility, 
$G(p^2=0) < \infty$, describes 'decoupling' characterized by a finite gluon 
propagator and a finite ghost dressing function at zero momenta. 
\cite{Boucaud:2008ji,Aguilar:2008xm,Fischer:2008uz}. Only the first of these
two possibilities, scaling, is in agreement with the Kugo-Ojima criterion
of well-defined global color charges \cite{Kugo,Lerche:2002ep}.

Nevertheless, both possibilities satisfy a confinement criterion developed
in Ref.~\cite{Braun:2007bx}: both types of infrared solutions lead to a 
confining Polyakov-loop potential. It has also been noted in Ref.~\cite{Braun:2007bx} 
that the confinement-deconfinement phase transition extracted from the
temperature dependence of the Polyakov-loop potential is not qualitatively  
affected by the choice of scaling or decoupling. Here we demonstrate that the
same is true for the deconfinement transition from the dual quark condensate.

To this end we note that the choice of the fit-function \Eq{gluefit} for the gluon
propagator and the ansatz for the quark-gluon vertex \Eq{vertexfit} correspond
to 'decoupling' in the deep infrared. An alternative choice
that fits the lattice results of Ref.~\cite{Cucchieri:2007ta} for the temperature
dependent gluon propagator equally well is given by
\beqa
Z_{T}^{scaling}(\vq,\oq,T) &=& Z_{T}(\vq,\oq,T) 
\left(\frac{q^2}{q^2 + \Lambda_{IR}^2}\right)^{2\kappa+1/2-1}\label{mod0} \\
Z_{L}^{scaling}(\vq,\oq,T) &=& Z_{L}(\vq,\oq,T)\,, \label{mod}
\eeqa
where $Z_{T,L}(\vq,\oq,T)$ are given by \Eq{gluefit}.
The effect of the extra factor is to convert the infrared behavior 
$Z_{T}(\vq,\oq,T) \sim q^2$ of the decoupling type of gluon dressing function 
to the scaling behavior $Z_{T}^{scaling}(\vq,\oq,T) \sim (q^2)^{2\kappa+1/2}$ with
the anomalous dimension $\kappa \approx 0.40$ \cite{Zwanziger:2001kw} of the 
three-dimensional theory. Corresponding scaling solutions from DSEs have been 
reported in Ref.~\cite{Cucchieri:2007ta}. Note that the electric part of the
gluon propagator always develops a screening mass and is therefore not modified
in \Eq{mod}.
The scale $\Lambda_{IR}$ is typically much smaller than $\Lambda_{QCD}$; here we
choose $\Lambda_{IR} = 100$ MeV in agreement with the results of \cite{Fischer:2008uz}.
The scaling of the quark-gluon vertex
in the quenched theory is also given by a power law \cite{Alkofer:2008tt}
and can be represented by the ansatz
\beq
\Gamma_\nu^{scaling}(q,k,p) = \Gamma_\nu(q,k,p) \left(\frac{q^2}{q^2 + \Lambda_{IR}^2}\right)^{-1/2-\kappa}\label{mod2}
\eeq
where $q$ is the gluon momentum and $\Gamma_\nu(q,k,p)$ is given by \Eq{vertexfit}.

Our numerical solutions for the quark condensate and the dressed Polyakov loop
are given in Fig.~\ref{scalvsdec}. The results with decoupling type of interaction already given in 
Fig.~\ref{res:torcont} in the main body of this work are shown together 
with the results obtained from the scaling type of interaction.
The differences are tiny. Below the transitions the effects are not visible.
Here the dynamically generated quark mass in the quark-DSE shields the
interaction in the deep infrared; the results are independent of the
behavior of the interaction below $\Lambda_{IR}$. We also find the same 
transition temperatures for both types of solutions. In the high temperature
phase where chiral symmetry is partly restored we observe small effects
in the ordinary condensate and almost no effect in the dual chiral condensate.
The main aspects of the two transitions are not concerned. Whether
there is some impact on the properties of quarks within the quark-gluon
plasma remains to be investigated.

\end{appendix}

\end{document}